\font\tenfrakturb=eufb10
\font\tenfraktur=eufm10
\font\tenmsbm=msbm10
\font\sevenfrakturb=eufb7
\font\sevenfraktur=eufm7
\font\sevenmsbm=msbm7
\font\fivefrakturb=eufb5
\font\fivefraktur=eufm5
\font\fivemsbm=msbm5
\newfam\bgothicfam
\newfam\gothicfam
\newfam\msbmfam
\textfont\bgothicfam = \tenfrakturb \scriptfont\bgothicfam=\sevenfrakturb
\scriptscriptfont\bgothicfam=\fivefrakturb

\textfont\gothicfam = \tenfraktur \scriptfont\gothicfam=\sevenfraktur
\scriptscriptfont\gothicfam=\fivefraktur

\textfont\msbmfam = \tenmsbm \scriptfont\msbmfam=\sevenmsbm
\scriptscriptfont\msbmfam=\fivemsbm

\def\frak{\tenfraktur\fam\gothicfam}
\def\Bbb{\tenmsbm\fam\msbmfam}

\catcode`@=11
\def\renewcounter#1{\@definecounter{#1}\@ifnextchar[{\@newctr{#1}}{}}
\documentstyle[12pt]{article}
\renewcounter{equation}[section]

\begin{document}
\title{Analytic Torsion on Hyperbolic Manifolds and the Semiclassical 
Approximation for Chern-Simons Theory}
\author{A.A. Bytsenko \thanks{E-mail: abyts@fisica.uel.br\,\,\,\,\,
On leave from Sankt-Petersburg State Technical University, Russia},\,\,\,
A.E. Gon\c calves \thanks{E-mail: goncalve@fisica.uel.br}\\
and\\
W. da Cruz \thanks{E-mail: wdacruz@fisica.uel.br}\\
Departamento de Fisica, Universidade Estadual de Londrina,\\
Caixa Postal 6001, Londrina-Parana, Brazil}


\maketitle
\begin{abstract}

The invariant integration method for Chern-Simons theory for gauge group 
$SU(2)$ and manifold $\Gamma\backslash H^3$ is verified in the semiclassical 
approximation. The semiclassical limit for the partition function associated 
with a connected sum of hyperbolic 3-manifolds is presented. We discuss
briefly $L^2-$ analytical and topological torsions of a manifold with 
boundary.

\end{abstract}

\section{Introduction}

It is known that topological invariants associated with $3$-manifolds can be
constructed within the framework of Chern-Simons gauge theory 
\cite{witt89-121-351}. These values were specified in terms of the axioms 
of topological quantum field theory \cite{moor89-123-77}, whereas equivalent 
derivation of invariants was also given combinatorially in
\cite{resh90-127-1,resh91-103-547}, where modular Hopf algebras related to 
quantum groups have been used. The Witten's (topological) invariants have been 
explicitly calculated for a number of 3-manifolds
and gauge groups
\cite{dijk90-129-393,kirb91-105-473,free91-141-79,jeff92-147-563,rama93-8-2285,
roza93u-99,roza94u-75,roza96-175-275}.

The semiclassical approximation for the Chern-Simons
partition function $Z_W(k)$ can be given by
the asymptotic $k\rightarrow\infty$ of Witten's invariant of a 
$3$-manifold $M$ and a gauge group $G$. Typically this expression is a 
partition function of quadratic functional. The asymptotic leads to a series of
$C^{\infty}-$ invariants associated to triplets $\{M;F;\xi\}$ with $M$ a smooth
homology 
$3-$ sphere, $F$ a homology class of framings of $M$, and $\xi$ an acyclic 
conjugacy class of ortogonal representations of fundamental group $\pi_1(M)$ 
\cite{axel94-39-173}. In addition the cohomology $H(M;Ad\,\xi)$ of $M$ 
with respect to the local system related to $Ad\,\xi$ vanishes. 
Description of an 
invariant of rational homology $3-$ spheres can be found in 
\cite{kont94,bott97u-01,bott98u-62}.

This note is an extension of the previous  paper \cite{byts97-505-641}. 
Our aim here will be to use the invariant integration method 
\cite{schw79-64-233,adam97u-47} in its simplest form 
for the semiclassical approximation in 
Chern-Simons theory. We do this analysing the partition function as well as 
the partition function related to a connected sum of hyperbolic 3-manifolds.
The resulting expressions are evaluated for gauge group $SU(2)$ and manifold 
$M=\Gamma\backslash H^3$, where
$H^3$ is the Lobachevsky space and $\Gamma$ is a co-compact discrete group
of isometries (see for detail \cite{byts96-266-1}).

In Sect. 2 we review the semiclassical approximation, involving the partition 
function and the canonical elliptic resolvent
of a quadratic functional. The explicit calculation of the Chern-Simons
partition function with gauge group $SU(2)$ is presented in Sect. 3. In 
Sect. 4 we discuss the $L^2-$ topological torsion of a compact manifold with
boundary. The semiclassical approximation for the partition function 
relating to a connected sum of hyperbolic $3-$ manifold is calculated in 
Sect. 5. Sect. 6 contains some remarks in summary.

\section{The Semiclassical Approximation for the Partition Function}

The partition function associated to Chern-Simons gauge
theory has the form
$$
Z_W(k)=\int {\cal D}A\exp[ikCS(A)]\mbox{,}\hspace{1.0cm} k\in {\Bbb Z}\mbox{,}
\eqno{(2.1)}
$$
where
$$
CS(A)=\frac{1}{4\pi}\int_{M}\mbox{Tr}\left(A\wedge dA+\frac{2}{3}A
\wedge A\wedge A\right)\mbox{.}
\eqno{(2.2)}
$$
The quantity $Z_W(k)$ is a (well-defined) topological invariant of $M$. 
The formal integration in (2.1) is over
the gauge fields $A$ in a trivial bundle, i.e. 1-forms on a 3-dimensional
manifold $M$ with values in Lie algebra ${\bf g}$ of gauge group $G=SU(N)$.

In the limit $k\rightarrow\infty$ Eq. (2.1) is given by its semiclassical
approximation, involving only partition functions of quadratic functionals 
\cite{witt89-121-351}:

$$
\sum_{[A_\theta]}\exp[ikCS(A_\theta)]\int {\cal D}B
\exp[ikCS_{A_\theta}^{(2)}(B)]
\mbox{,}
\eqno{(2.3)}
$$
$$
CS_{A_\theta}^{(2)}(B)=\frac{1}{4\pi}\int_{M}
\mbox{Tr}(B\wedge d^{A_\theta}B)
\mbox{.}
\eqno{(2.4)}
$$
In Eq. (2.3) the sum is taken over representatives $A_\theta$ for each point
$[A_\theta]$ in the moduli-space of flat gauge fields on $M$. In addition, 
$B$ is Lie-algebra-valued 1-form and $d^{A_\theta}$ is the covariant 
derivative
determined by $A_\theta$, namely $d^{A_\theta}B=dB+[A_\theta,B]$.
We shall use the invariant integration method \cite{schw79-64-233,adam97u-47},
which enables the partition functions in  Eq. (2.3) to
be evaluated in complete generality. 

Let $X$ be a compact oriented Riemannian manifold without boundary, and
${\rm dim}(X) = (2m+1)$ is the dimension of manifold.
A quadratic functional (like functional (2.4)) can be defined on the space 
${\cal G}={\cal G}(X,\xi)$ 
of smooth sections in a real Hermitian vector bundle $\xi$ over $X$. 
Let $D_q$ denote the restriction of a flat connection map $D$ 
on the space $\Omega(X,\xi)$ of differential forms on $X$ with values in
$\xi$ to the space
$\Omega^q(X,\xi)$ of $q$-forms and $H^q(X;\xi)=H^q(D)=
\mbox{ker}(D_q)\left[\Im(D_{q-1})\right]^{-1}$ are the cohomology spaces. 
One can construct the inner products $\langle\cdot\,,\cdot\rangle_m$ in the 
space $\Omega^m(X,\xi)$ and the quadratic functional on this space,
$S_D(B)=\langle B,*DB\rangle_m$,\,\, $(*)$ is the Hodge-star map. 
A canonical topological elliptic resolvent $R(S_D)$ for the quadratic 
functional can be written as follows:

$$
0\stackrel{0}{\longmapsto}\Omega^0(M,\xi)\stackrel{D_0}{\longmapsto}...
\stackrel{D_{m-2}}{\longmapsto}\Omega^{m-1}(M,\xi)\stackrel{D_{m-1}}
{\longmapsto}\mbox{ker}(S_D)\stackrel{0}{\longmapsto}0\mbox{.}
\eqno{(2.5)}
$$
From Eq. (2.5) it follows that for the resolvent $R(S_D)$ we have 
${\cal G}_q=\Omega^{m-q}(X,\xi)$ and $H^q(R(S_D))=H^{m-q}(D)$. Note that 
$S\geq 0$ and therefore $\mbox{ker}(S_D)\equiv \mbox{ker}(D_m)$.

Let $X=M$ be a closed hyperbolic 3-manifold, $G$  a compact simple Lie group,
${\bf g}$ the Lie algebra of $G$. The inner products in the
space $\Omega^q(M,{\bf g})$ of ${\bf g}-$ valued q-forms naturally can be
chosen as $\langle B,*d_1^{A_\theta}B\rangle$,\,\,$D_1\equiv*d_1^{A_\theta}$. 
Thus the canonical elliptic resolvent for quadratic functional (2.4) takes the 
form:
$$
0\longmapsto\Omega^0(M,{\rm g})\stackrel{d_0^{A_\theta}}{\longmapsto}
\mbox{ker}(d_1^{A_\theta})\longmapsto 0\mbox{,}
\eqno{(2.6)}
$$
and the resolvent has cohomology spaces $H^0(R(S))=H^1(d^{A_\theta}),\,\,\,
H^1(R(S))=H^0(d^{A_\theta})$.

\section{Gauge Group $G=SU(2)$}

The invariant integration method \cite{schw79-64-233} leads to the following
expression for the Chern-Simons partition function 
\cite{adam95u-95,adam97u-47}:

$$
Z_{sc}(k)=\int_{{\cal M}}{\cal D}[A_{\theta}]V_{\cal G}(H_{A_{\theta}})^{-1}
e^{i\left(\frac{\pi}{4}\eta(A_{\theta})+kS(A_{\theta})\right)}
\left(\frac{k\lambda}{4\pi^2}\right)^{-\zeta(A_{\theta})/2}
[T_{an}(M)]^{1/2}
\mbox{,}
\eqno{(3.1)}
$$
where $T_{an}(M)$ is the Ray-Singer analytic torsion of $D_1$ 
\cite{ray71-7-145}, $\eta(A_{\theta})$ and $\zeta(A_{\theta})$ are the analytic
continuations to $s=0$ of the eta and the zeta functions respectively 
\cite{eliz94,byts96-266-1},
$\zeta(A_{\theta})={\rm dim}H^0(A_{\theta})-{\rm dim}H^1(A_{\theta})$
and $H^q(A_{\theta})$ the $q-$th cohomology space of $d^{A_{\theta}}$.

Let $G=SU(2)$ and $M=\Gamma\backslash H^3$. A convenient choise of orthogonal
basis (determing a left invariant metric on $SU(2)$) for 
${\bf g}=su(2)$ is
$$
a_1=\frac{1}{2}\left(\begin{array}{ll}
0\,\,\,i\\
i\,\,\,0
\end{array}\right),\,\,\,\,\,\,\,
a_2=\frac{1}{2}\left(\begin{array}{ll}
\,\,\,0\,\,\,\,\,1\\
-1\,\,\,0
\end{array}\right),\,\,\,\,\,\,\,
a_3=\frac{1}{2}\left(\begin{array}{ll}
i\,\,\,\,\,\,\,\,\,\,\,0\\
0\,\,\,-i
\end{array}\right)
\mbox{,}
\eqno{(3.2)}
$$
then it follows $\lambda=1/2$. We define $U(1)\in SU(2)$ by \cite{adam97u-47}
$$
U(1)\stackrel{def}{=}\left\{e^{a_3\theta}=\left(\begin{array}{ll}
e^{i\frac{\theta}{2}}\,\,\,0\\
0\,\,\,e^{-i\frac{\theta}{2}}\end{array}\right)
|\theta\in[0,4\pi[\right\}
\mbox{.}
\eqno{(3.3)}
$$
For any $U(1)$ flat connection on a manifold with Betti number 
$b_1(M)=0$ we have

$$
H_{A_{\theta}}=U(1),\,\,\,{\rm dim}H_{A_{\theta}}^0={\rm dim}H_{A_{\theta}}=1,
\,\,\,{\rm dim}H_{A_{\theta}}^1=0
\mbox{,}
\eqno{(3.4)}
$$
while for trivial $SU(2)$ connection,
$$
H_{A_{\theta}}=SU(2),\,\,\,{\rm dim}H_{A_{\theta}}^0={\rm dim}H_{A_{\theta}}=3,
\,\,\,{\rm dim}H_{A_{\theta}}^1=0
\mbox{,}
\eqno{(3.5)}
$$ 
and $\zeta(0)=3-0=3$. The volumes of $U(1)$ and $SU(2)$ are equal to
$V(U(1))=4\pi$ and $V(SU(2))=16\pi^2$ respectively. Thus we get

$$
Z_{sc}(k)=\int_{{\cal M}}{\cal D}\theta V(SU(2))^{-1}
e^{i\pi kS(0)/4}\left(\frac{k\lambda}{4\pi^2}\right)^{-\zeta(0)/2}
\left[T_{an}(M)\right]^{1/2}
$$
$$
=\sqrt{2}\pi k^{-3/2}[T_{an}(M)]^{1/2}
\mbox{.}
\eqno{(3.6)}
$$

Using the Hodge decomposition, the cohomology $H(X;\xi)$ can be embedded into
$\Omega(X;\xi)$ as the space of harmonic forms. This embedding induces a
norm $|\cdot|^{RS}$ on the determinant line ${\rm det}H(X;\xi)$. The
Ray-Singer norm $||\cdot||^{RS}$ on ${\rm det}H(X;\xi)$ is defined by 
\cite{ray71-7-145}

$$
||\cdot||^{RS}\stackrel{def}=|\cdot|\prod_{q=0}^{{\rm dim}X}
\left[\exp\left(-\frac{d}{ds}
\zeta_q(s)|_{s=0}\right)\right]^{(-1)^qq/2}
\mbox{,}
\eqno{(3.7)}
$$
where the zeta function $\zeta_q(s)$ of the Laplacian acting on the space of
$q-$ forms orthogonal to the harmonic forms has been used. For a closed
connected orientable smooth manifold of odd dimension and for Euler structure
$\eta\in {\rm Eul}(X)$ the Ray-Singer norm of its cohomological torsion
$T_{an}(X;\eta)=T_{an}(X)\in {\rm det}H(X;\xi)$ is equal to the positive
square root of the absolute value of the monodromy of $\xi$ along the 
characteristic class $c(\eta)\in H^1(X)$ \cite{farb98u-137}: 
$||T_{an}(X)||^{RS}=|{\rm det}_{\xi}c(\eta)|^{1/2}$. In the special case where
the flat bundle $\xi$ is acyclic $(H^q(X;\xi)=0)$ we have
$$
\left[T_{an}(X)\right]^2
=|{\rm det}_{\xi}c(\eta)|
\prod_{q=0}^{{\rm dim}X}\left[\exp\left(-\frac{d}{ds}
\zeta_q(s)|_{s=0}\right)\right]^{(-1)^{q+1}q}
\mbox{.}
\eqno{(3.8)}
$$
For odd-dimensional manifold the Ray-Singer norm is topological invariant: it
does not depend on the choice of metric on $X$ and $\xi$, used in the
construction. But for even-dimensional $X$ this is not the case
\cite{bism92}.

\section{Manifold with Boundary}

The topological Chern-Simons action on a manifold with boundary is    

$$
CS(A)=\frac{1}{4\pi}\int_{M}\mbox{Tr}\left(A\wedge dA+\frac{2}{3}A
\wedge A\wedge A\right)
+\frac{1}{4\pi}\int_{\partial M}{\rm Tr}A_zA_{\bar{z}}
\mbox{.}
\eqno{(4.1)}
$$
For closed manifold the boundary term in Eq. (4.1) disappears and the 
integrand in Eq. (2.1), $\exp(ikCS(A))$, is gauge invariant. This invariance
is broken for manifold with boundary since under the decomposition
$A={\rm g}^{-1}d{\rm g}+{\rm g}^{-1}\tilde{A}{\rm g}$ the action becomes 
\cite{ogur89-229-61,carl91-362-111}: $CS(A)=CS(\tilde{A})+I_{WZW}({\rm g},\tilde{A}_z)$,
where the action $I_{WZW}({\rm g},\tilde{A}_z)$ of a chiral WZW model on the
boundary $\partial M$ is given by
$$
I_{WZW}({\rm g},\tilde{A}_z)=\frac{1}{4\pi}\int_{\partial M}\mbox{Tr}
\left({\rm g}^{-1}\partial_z{\rm g}{\rm g}^{-1}\partial_{\bar{z}}{\rm g}
-2{\rm g}^{-1}\partial_{\bar{z}}{\rm g}\tilde{A}_z\right)
+\frac{1}{12\pi}\int_{M}{\rm Tr}\left({\rm g}^{-1}d{\rm g}\right)^3
\mbox{.}
\eqno{(4.2)}
$$
The gravitational analog of the WZW action is especially  difficult to 
construct. Gravity is not quite a pure Chern-Simons theory, however 
for Euclidean gravity in three spacetime dimensions with negative cosmological
constant $\Lambda=-1$ 
one can define an $SL(2,\Bbb C)$ gauge field $A^a\stackrel{def}=
\omega^a+ie^a,\,\,\, \bar{A}^a\stackrel{def}=\omega^a-ie^a$, 
where $e^a=e_{\mu}^adx^{\mu}$ is a triad and 
$\omega^a=(1/2)\epsilon^{abc}\omega_{\mu bc}dx^{\mu}$ is a spin connection.
The standard Einstein action then is easily found to be  
$$
I_{GR}=CS(A)-CS(\bar{A})
\mbox{.}
\eqno{(4.3)}
$$ 

It has been shown however \cite{moor89-220-422,elit89-326-108} that the 
covariant Chern-Simons action supplemented
with boundary conditions $A_{\bar{z}}=\bar{A}_z=0$ does not require any
boundary terms. Nevertheless, in this case the analytic torsion of a 
Riemannian compact manifold with boundary has to be computed.
We recall here the definition of $L^2-$ analytic torsion 
(see also \cite{lott92-35-471,math92-107-369}). Let ${\cal X}\mapsto X$ be 
the universal covering of a compact connected Riemannian manifold $X$, which
is of determinant-class. Define the $L^2-$ analytic torsion of $X$ by

$$
T_{an}^{(2)}(X)\stackrel{def}{=}\sum_{q\geq 0}(-1)^qq\left[
\frac{d}{ds}\frac{1}{\Gamma(s)}\int_{0}^1t^{s-1}{\rm Tr}_{\Gamma}e^
{-t\triangle_q^{\perp}[\cal X]}dt|_{s=0}\\ \right.
$$
$$
\left. +\int_{1}^{\infty}{\rm Tr}_{\Gamma}e^
{-t\triangle_q^{\perp}[\cal X]}\frac{dt}{t}\right]
\mbox{.}
\eqno{(4.4)}
$$
Here $\triangle_q[\cal X]$ is the Laplacian on $q-$ forms on the universal
covering ${\cal X}$ considered as an unbounded self-adjoint operator,
$\triangle_q^{\perp}[\cal X]$ is the operator from the orthogonal complement
of the kernel of $\triangle_q[\cal X]$ to itself which is obtained from
$\triangle_q[\cal X]$ by restriction, and
${\rm Tr}_{\Gamma}\exp(-t\triangle_q^{\perp}[\cal X])$ is the normalized trace for the $\Gamma=\pi_1(X)-$ action (see for detail \cite{luck97u-09}). The first
integral in Eq. (4.4) is defined for $\Re s>({\rm dim} X)/2$, but it has a
meromorphic extension to the complex plane $\Bbb C$ with no pole in $s=0$.
The second integral converges since $X$ is of determinant-class by
assumption. Let $\{{\rm g}_u\}_{0\leq u\leq 1}$ be a smooth family of
Riemannian metric on $X$. Let set $V_q=V_q({\rm g}_u)\stackrel{def}{=}
\left[\frac{d}{du}{*}_q({\rm g}_u)\right]{*}_q({\rm g}_u)^{-1}$. It can be
shown \cite{chee79-109-259,luck97u-09} that analytic torsion $T_{an}(X)=
T_{an}(X;{\rm g}_u)$ and $L^2-$ analytic torsion $T_{an}^{(2)}(X)=
T_{an}^{(2)}(X;{\rm g}_u)$ are smooth function of $u$, whereas

$$
\frac{d}{du}\left[T_{an}^{(2)}(X)- T_{an}(X)\right]_{u=0}=
\sum_{q}(-1)^{q+1}\left({\rm Tr}_{\Gamma}V_q|_{{\rm ker}\triangle[\cal X]}
-{\rm Tr}_{\Gamma}V_q|_{{\rm ker}\triangle[X]}\right)
\mbox{,}
\eqno{(4.5)}
$$
$$
\frac{d}{du}[T_{an}^{(2)}(X)]_{u=0}=\sum_{q}(-1)^{q}\left(d_q-
{\rm Tr}_{\Gamma}V_q|_{{\rm ker}\triangle[\cal X]}\right)
\mbox{.}
\eqno{(4.6)}
$$

Let $\bar{X}$ be a compact connected manifold with boundary whose interior
$X$ comes with a complete hyperbolic metric of finite volume. 
It can be shown
\cite{luck97u-09} that the $L^2$- topological torsion of
$\bar{X}$ and the $L^2$-analytical torsion of Riemannian manifold $X$
are equal. The $L^2$- topological torsion $T_{top}^{(2)}(\bar{M})$ in 
dimension 3 is proportional to the hyperbolic volume of $M$, with a constant
of proportionality which depends only on the dimension 
\cite{luck97u-09,lott95-120-15}. This gives a complete calculation of the 
$L^2$- topological
torsion of a compact $L^2$- acyclic 3-manifolds which admit a geometric JSJT-
decomposition, i.e. the decomposition of Jaco-Shalen and Johannson by a 
minimal family of pairwise non-isotopic incompressible not boundary-parallel
embedded 2-tori into Seifert pieces and atoroidal pieces. Moreover, there is
a dimension constant $C_N$ \cite{luck97u-09} such that
$T_{an}^{(2)}(X)=C_N\cdot{\rm vol}(M)$, and $C_N=0$ for $N$ even, 
$C_3=-1/(3\pi), C_5=-3/(\pi^2), C_7=11/(2\pi^3)$.

\section{Connected Sum of 3-Manifolds}

The analytic torsion $T_{an}(\Gamma\backslash H^3)$ can be expressed in terms 
of the Selberg zeta functions ${\cal Z}_p(s)$. Let the flat bundle $\xi$ is
acyclic. The Ruell's zeta 
function in three dimension associated with closed oriented hyperbolic 
manifold $M=\Gamma \backslash H^3$ has the form

$$
{\cal R}_{\chi}(s)=\prod_{p=0}^2{\cal Z}_p(p+s)^{(-1)^p}=\frac{{\cal Z}_0(s)
{\cal Z}_2(2+s)}{{\cal Z}_1(1+s)}
\mbox{.}
\eqno{(5.1)}
$$
The function ${\cal R}_{\chi}(s)$ extends meromorphically to the entire 
complex plane $\Bbb C$ \cite{deit89-59-101}. For the Ray-Singer torsion one 
gets \cite{byts97-505-641}
$$
[T_{an}(\Gamma\backslash H^3)]^2={\cal R}_{\chi}(0)
=\frac{[{\cal Z}_0(2)]^2}{{\cal Z}_1(1)}
\exp\left(-\frac{V({\cal F})}{3\pi}\right)
\mbox{.}
\eqno{(5.2)}
$$
In the presence of non-vanishing Betti numbers $b_j=b_j(M)$ we have 

$$
[T_{an}(\Gamma\backslash H^3)]^2=\frac{(b_1-b_0)!
[{\cal Z}_0^{(b_0)}(2)]^2}
{[b_0!]^2 {\cal Z}_1^{(b_1-b_0)}(1)}
\exp\left(-\frac{V({\cal F})}{3\pi}\right)
\mbox{,}
\eqno{(5.3)}
$$
and using Eq. (3.6) one obtains,
$$
Z_{sc}(k)=\sqrt{2}\pi k^{-\frac{3}{2}}\left[
\frac{(b_1-b_0)!({\cal Z}_0^{(b_0)}(2))^2}
{(b_0!)^2{\cal Z}_1^{(b_1-b_0)}(1)}\right]^{\frac{1}{4}}
\exp\left(-\frac{V({\cal F})}{12\pi}\right)
\mbox{.}
\eqno{(5.4)}
$$

In Chern-Simons theory the partition function for a connected sum 
${\frak M}=M_1$\# $M_2$\# ... \# $M_N$ can be written as follows 
\cite{witt89-121-351}
$$
Z({\frak M})=\frac{\bigotimes_{\ell=1}^NZ(M_\ell)}{[Z(S^3)]^{N-1}}
\mbox{.}
\eqno{(5.5)}
$$
The fundamental group for 3-sphere $\pi_1(S^3)$
is trivial and $M$ consists of a single point corresponding to 
$A_{\theta}$. Since the Ray-Singer torsion is to be equal one (cf. 
\cite{adam97u-47}), using Eq. (5.1)
we get $Z_{sc}(k)=\sqrt{2}\pi k^{-3/2}$. The partition function associated 
with the semiclassical approximation then takes the form
$$
Z_{sc}({\frak M})=\left(\frac{k^3}{2\pi^2}\right)^{\frac{N-1}{2}}
\bigotimes_{\ell=1}^N
Z_{sc}(M_{\ell})=\sqrt{2}\pi k^{-\frac{3}{2}}\bigotimes_{\ell=1}^N
|{\cal R}_{\chi_{(\ell)}}(0)|^{\frac{1}{2}}
\mbox{,}
\eqno{(5.6)}
$$
while in the presence of non-vanishing Betti numbers $b_{j\ell}=b_j(M_{\ell})$
one gets
$$
Z_{sc}({\frak M})=\sqrt{2}\pi k^{-\frac{3}{2}}
\bigotimes_{\ell=1}^N
\left[\frac{(b_{1\ell}-b_{0\ell})!({\cal Z}_0^{(b_{0\ell})}(2))^2}
{(b_{0\ell}!)^2 {\cal Z}_1^{(b_{1\ell}-b_{0\ell})}(1)}
\right]^{\frac{1}{4}}
\exp\left[-\frac{1}{12\pi}
\bigoplus_{\ell=1}^NV({\cal F}_{\ell})\right]
\mbox{.}
\eqno{(5.7)}
$$
In the case of non-trivial characters $b_0(M_{\ell})=0$. If $b_1=0$ then 
Eq. (5.2) holds.

For trivial character one has $b_0=1$ 
(for any closed manifold) and $b_1=0$ for an infinite number of 
$M=\Gamma\backslash H^3$. The function ${\cal R}(s)$ 
has a zero at $s=0$ of order $4$ \cite{frie86-84-523}. 
However, there is a class of compact sufficiently large hyperbolic manifolds 
which admit arbitrarly large value of $b_1(M)$. 
Sufficiently large manifold $X$ contains a surface $S$ whereas 
$\pi_1(S)$ is infinite and $\pi_1(S)\subset \pi_1(X)$. In general, hyperbolic 
manifolds have not been completely classified and therefore a systematic 
computation is not yet possible. However it is not the case for sufficiently
large manifolds, classification of which we mention here. These manifolds give 
an essential contribution to the partition functions (5.4) and (5.7).

It is known that any 3-manifold can be triangulated, 
and hence can be partioned into handles. 
Since the existence and uniqueness of a decomposition of an orientable 
manifold as the sum of simple orientable parts have been established 
(see \cite{miln62-84-1,hemp76}), the question of homeomorpy can be 
considered only for irreducible manifolds. 
The method proposed by Haken \cite{hake61-105-245}
permits to describe all normal surfaces of a 3-manifold $X$ which has been 
partioned on handles previously. The Haken's theory of normal surfaces was 
further verified for the procedure of geometric summation of surfaces. 
As a result the classification theorem \cite{matv97-52-147} says: 
there exists an algorithm for enumerating of all the Haken manifolds 
and there exists an algorithm for recognizing homeomorphy of the 
Haken manifolds.

\section{Conclusions}

We have derived explicit formulae for the semiclassical approximation for the
Chern-Simons partition functions, using the invariant integration method. The
resulting expressions are explicitly evaluated for gauge group $SU(2)$ and
manifold $\Gamma\backslash H^3$. Manifolds with boundary and $L^2-$ topological
torsion of such manifolds has been discussed briefly. It will be 
interesting to obtain the exact formulae for the partition function (and for
the partition function of connected sum of manifolds) in the large $k-$ limit
for more complicate situations, e.g. when a discrete group $\Gamma$ contains
elliptic or parabolic elements, and for gauge groups other than $SU(2)$.
We hope that proposed discussion of the topological invariants will be 
interesting in view of future applications to concrete problems in quantum
field theory.

\section{Acknowledgements}

We thank Profs. L. Vanzo and S. Zerbini for useful discussions.
A.A. Bytsenko wishes to thank CNPq and the Department of Physics of
Londrina University for financial support and kind hospitality. The research
of A.A. Bytsenko was supported in part by Russian Foundation for Basic
Research (grant No. 98-02-18380-a) and by GRACENAS (grant No. 6-18-1997).

\end{document}